\documentclass[a4paper,reqno,10pt]{amsart}

\usepackage{setspace}
\usepackage{hyperref}
\usepackage{amsmath}
\usepackage{amsthm}
\usepackage{amssymb}
\usepackage{amsfonts}
\usepackage{mathrsfs}
\usepackage{mathtools}
\usepackage{graphicx} 
\usepackage{cite}
\usepackage{hyperref} 
\usepackage{eucal}
\usepackage{accents}
\numberwithin{equation}{section}
\usepackage{xcolor}

\newcommand{\timebullet}{\hbox{.}}
\newcommand\bdot[1]{\accentset{\halfspace\quart\timebullet}{#1}}

\newcommand\TG{\stackrel{\hbox{${}_{{}^{{}_\text{TG}}}$}}{=}}

\newcommand{\za}{\bar a}  
\newcommand{\zb}{\bar b}
\newcommand\zc{\bar c}  \newcommand\zd{\bar d}
\newcommand\ze{\bar e}

\newcommand\pr{\text{\textsf{pr}\,}}

\newcommand{\LRm}{\tilde\Lscr_\text{R}}

\newcommand\bfE{\mathbf{E}}
\newcommand\bfB{\mathbf{B}}
\newcommand\bfA{\mathbf{A}}
\newcommand\bfC{\mathbf{C}}

\newcommand{\ms}[1]{\scriptscriptstyle{#1}}

\newcommand{\eps}{\epsilon} 

\usepackage{color}

\newcounter{mnotecount}[section]

\theoremstyle{plain}
\newtheorem{theorem}{Theorem}[section]

\newtheorem{remark}[theorem]{Remark}

\newcommand{\pd}{\partial}
\newcommand{\lrpd}{\halfspace \overset{\quart{}_{\ms\leftrightarrow}}
           {\nhalf\pd}\!\nhalf\nquart}

\newcommand{\dual}[1]{{}^\star\!#1}

\newcommand\Ascr{{\mathcal A}} 
\newcommand\Fscr{{\mathcal F}} 
\newcommand{\Hscr}{\mathcal H}
  
\newcommand\Lscr{{\mathcal L}}
\newcommand\Mscr{{\mathcal M}} \newcommand\Nscr{{\mathcal N}}

\newcommand\Xscr{{\mathcal X}} 
\newcommand\Uscr{{\mathcal U}}

\newcommand\Yscr{{\mathscr Y}}

\newcommand\barAscr{\hbox{$\bar{\phantom{\Ascr}}\!\!\!\!\!\Ascr$}}
\newcommand\barFscr{{\bar{\Fscr}}}

\newcommand\Esf{\mathsf{E}} 

\renewcommand{\div}{\mathop{\rm div}\nolimits}

\newcommand{\halfspace}{\kern.083333em}   
\newcommand{\quart}{\kern.0416675em}  
\newcommand{\nhalf}{\kern-.083333em}   
\newcommand{\nquart}{\kern-.0416675em}  

\title{Helicity, spin, and infra-zilch of light: \\a Lorentz covariant formulation}

\author[S. Aghapour]{Sajad Aghapour}
\address[S. Aghapour]{School of Physics, Institute for Research in Fundamental Sciences (IPM), Tehran, Iran \and Albert Einstein Institute, Am M\"uhlenberg 1, D-14476 Potsdam, Germany}
\email{aghapour@ipm.ir}

\author[L. Andersson]{Lars Andersson}
\address[L. Andersson]{Albert Einstein Institute, Am M\"uhlenberg 1, D-14476 Potsdam, Germany }
\email{laan@aei.mpg.de}

\author[K. Rosquist]{Kjell Rosquist}
\address[K. Rosquist]{AlbaNova University Center, Stockholm University, Department of Physics, SE-106 91 Stockholm, Sweden}
\email{kr@fysik.su.se}

\begin{document}

\begin{abstract}
 In this paper, a novel conserved  Lorentz covariant tensor, termed the helicity tensor,  is introduced in Maxwell theory. The conservation of the helicity tensor expresses the conservation laws contained in the helicity array, introduced by Cameron et al. \cite{Cameron_etal:2012}, including helicity, spin, and the spin-flux or infra-zilch. The Lorentz covariance of the helicity tensor is in contrast to previous formulations of the helicity hierarchy of conservation laws, which required the non-Lorentz covariant transverse gauge. The helicity tensor is shown to arise as a Noether current for a variational symmetry of a duality-symmetric Lagrangian for Maxwell theory.
This symmetry transformation generalizes the duality symmetry and includes the symmetry underlying the conservation of the spin part of the angular momentum.

\end{abstract}

\date{\today} 

\maketitle

\section{Introduction}
The physical relevance of the spin and orbital parts of the angular momentum of the Maxwell field was first demonstrated by 
Allen et al.\ \cite{1992PhRvA..45.8185A} and van Enk and Nijenhuis \cite{1994JMOp...41..963V} in the early 1990's. 
These and subsequent works have led to the understanding that these two parts of angular momentum are separately conserved. The related conservation laws have recently been re-examined by several groups, including I.\ and Z.\  Bialynicki-Birula  \cite{Bialynicki-Birula2011}, Bliokh et al.\ \cite{Bliokh_2014} and Barnett et al.\ \cite{Barnett_2016}. The two parts of the angular momentum are physically meaningful observable quantities despite the fact that they do not satisfy the commutation relations of angular momenta. 
Among the phenomena related to the just mentioned conservation laws is the experimentally verified spin Hall effect of light, cf. Bliokh et al. \cite{Bliokh2015} and references therein. 

The notion of electromagnetic (or optical) helicity was introduced by Trueba and Ra{\~{n}}ada based on similar notions defined in the context of magnetohydrodynamics and fluid dynamics \cite{Trueba_1996}. The conservation of helicity is due to the fact that it is the density of a Noether current associated to the duality symmetry of the free Maxwell equations, while the spin part of the angular momentum turns out to be the flux of helicity \cite{Deser1976,Cameron_etal:2012}. The electromagnetic helicity has been used in describing the interaction of light beams with chiral molecules and nano-particles \cite{Jungho2020}. These quantities also generalize to linearized gravitational fields and acoustic waves, cf. e.g. \cite{PhysRevB.99.174310,Burns_2020,Barnett2014,Andersson_etal:2018}.

The zilch tensor, a third rank tensor constructed from the product of the Maxwell field and its first derivative was introduced by Lipkin \cite{Lipkin:1964}, see also Morgan \cite{Morgan:1964} and Kibble \cite{Kibble:1965}. The zilch tensor is conserved for the Maxwell field in the absence of sources. The hierarchy of conservation laws expressed by the zilch tensor, like the above mentioned helicity and related conservation laws, constitute new conserved quantities, that are not equivalent to the classical conservation laws related to Poincar\'e and conformal symmetries of Minkowski space, cf.\ \cite{MR1885280} and references therein. 

The zilch conservation laws were generalized by Morgan to a hierarchy of quadratic tensors with an arbitrary number of derivatives. 
Morgan also found another such hierarchy generalizing the Maxwell stress-energy tensor.
It has been proposed that the purely temporal component of the zilch tensor describes the chirality of the electromagnetic fields \cite{Tang2010}. Electromagnetic helicity and chirality are proportional for monochromatic waves but are in general distinct notions. It is still under debate which one of these quantities is more appropriate for describing relevant properties of light in its interaction with chiral matter, cf. \cite{Cameron2015} for discussion.

Motivated by the structure of the zilch tensor, the conserved currents associated to the 
helicity density $\Hscr$, spin density $S_a$, and infra-zilch $\Sigma_{ab}$ have been arranged by Cameron et al.\ \cite{Cameron_etal:2012} in what they named the helicity array, $\Nscr^{abc}$. Here, the indices label the spacetime components of the array, taking the values $\{0,1,2,3\}$.
The  array is by construction symmetric in the first two indices, 
$\Nscr^{abc} = \Nscr^{(ab)c}$. The corresponding conservation laws could then be represented in the compact form $\pd_c\Nscr^{abc} = 0$, valid provided the field equations hold and the transverse gauge condition is imposed. However, as stated in 
\cite{Cameron_etal:2012}, the array $\Nscr^{abc}$ is not a tensor. Further, the analysis in  \cite{Cameron_etal:2012} as well as earlier papers on the subject including the papers cited above makes use of the transverse gauge condition, i.e.\ a combination of temporal gauge $A_0 = 0$ and Coloumb gauge. Due to this elliptic gauge condition, the helicity array is neither tensorial nor Lorentz covariant.

As was noted in \cite{Cameron_etal:2012}, the structure of the helicity array is closely related to the zilch tensor but differs in the number of derivatives. As mentioned above, the zilch tensor is of the third order in derivatives of the Maxwell potential, while the helicity array is of the first order. The similarities in the structures is illustrated by the formal transformation
$(\bfA, \bfC) \rightarrow (\nabla\times \bfA, \nabla\times \bfC)$
of the electric and magnetic potentials mapping the helicity array to the zilch tensor, $\Nscr_{abc} \rightarrow Z_{abc}$, cf. \cite{Cameron_etal:2012}. Here $\bfA, \bfC$ are the 3-vector parts of the electric and magnetic potentials for the Maxwell field. The fact that this mapping involves both potentials is an indication that a tensor description of the helicity array is related to a self-dual formulation of Maxwell theory.

Let $\Ascr_a = A_a + i\halfspace C_a$
 be the complex potential formed from the electric and magnetic potentials $A_a, C_a$, and let 
\begin{align} \label{eq:Yscr}
    \Yscr_{abc} = \tfrac{i}2\,\bar{\Ascr}_a \lrpd_{c} \,\Ascr_b\ .
\end{align}
where $\bar{\Ascr}_a$ denotes complex conjugation, and we have used the bi\-directio\-nal-arrow derivative notation, which is often used in the quantum theory literature. This is defined in the present case by $\bar{\Ascr}_a \lrpd_{c} \,\Ascr_b\  = \bar \Ascr_a \pd_c \Ascr_b - \Ascr_b \pd_c \bar \Ascr_a$ and extends to general fields in the same manner. 
In section \ref{sec:Htensor} we shall define the helicity tensor by 
\begin{align}\label{helicity_tensor1}
    \Hscr_{abc} =(\hat P \Yscr)_{abc},
\end{align}
where $\hat P$ is a linear involution, which transforms the components of the auxiliary tensor $\Yscr_{abc}$ with even parity to odd parity.
By this construction, $\Hscr_{abc}$ is a real tensor, which like the helicity array, and the classical angular momenta, has odd parity. The helicity, spin and infra-zilch conservation laws are consequences of 
\begin{align} 
\partial_c \Hscr^{abc} = 0,
\end{align} 
which is valid provided that the vacuum Maxwell equations $\partial_b \Fscr_a{}^b = 0$, and the Lorenz gauge condition $\partial_a \Ascr^a = 0$ hold, where $\Fscr_{ab} = \pd_a \Ascr_b - \pd_b \Ascr_a$  is the complex field strength. 

The helicity tensor $\Hscr_{abc}$ is a quadratic tensor constructed from the electric and magnetic potentials and their first derivatives and contains the same information as that contained in the components of the helicity array. For details, see section \ref{sec:Htensor}.
We further show in section \ref{sec:Noether} that the helicity tensor arises as the Noether current of a symmetry of the gauge extended, duality-symmetric Maxwell action.

\subsection*{Overview of this paper}
In section \ref{sec:pre}, we review the construction of helicity array and the interrelated conservation laws of its components. Then, in section \ref{sec:Htensor}, we introduce the covariant helicity tensor. It is a Lorentz covariant tensor that is shown to contain the same information as the helicity array. 
In this analysis we make use of the $1+3$ decompositions discussed in appendix \ref{appendix2}, in terms of the notation introduced in appendix \ref{appendix}.
In section \ref{sec:Noether}, we demonstrate that the helicity tensor arises as a Noether current for a symmetry of the duality-symmetric action for the Maxwell field, and find the generator of the corresponding symmetry. In particular this is a symmetry of the duality-symmetric Lagrangian with  Lorenz gauge fixing terms. 

\section{The helicity array in Maxwell theory}\label{sec:pre}

We shall consider fields on Minkowski spacetime with metric $g_{ab}$. We use the index notation with indices $a, b, c ,\cdots$ taking values $0,\cdots,3$, using $0$ for a temporal index and overlines indices $\za, \zb, \dots$ for spatial indices, cf. appendix \ref{appendix}. 
We shall sometimes use the standard 3-dimensional vector notation, with bold-face capital letters $\mathbf A, \dots$ denoting 3-vectors. In this context, $\div$ and $\times$ denote the divergence and cross-product. Note that we also use the notation $\div$ for the 4-dimensional divergence.

In this section we shall review the helicity array and the associated conservation laws. 
Let $A_a, C_a$ be the electric and magnetic potentials and let $F_{ab} = \pd_a A_b - \pd_b A_a$, $G_{ab} = \pd_a C_b - \pd_b C_a$ be the field strength and its dual. In this section, unless otherwise stated, we shall use the transverse gauge condition
\begin{equation}\label{transverse_gauge}
   A_0 = 0  = C_0 \,,\quad \div\mathbf A  = 0 = \div\mathbf C \,.
\end{equation}

The helicity current 
\begin{equation}\label{Hcurrentvec}
   J^a = \tfrac12 (G^{ab} A_b - F^{ab} C_b) 
\end{equation}
satisfies the conservation law\footnote{\label{foot:cong} Here and below $\cong$ signifies equality modulo the field equations $F^b{}_{a,b}=0$ and $G^b{}_{a,b}=0$.}
\begin{equation}\label{Jcons}
   \pd_a J^a = \tfrac12 A^a G^b{}_{a,b} - \tfrac12 C^a F^b{}_{a,b} \cong 0  \,.
\end{equation}
The temporal part of the helicity current is referred to as the helicity density and has the form \cite{Cameron_etal:2012} 
\begin{equation}
   \Hscr = J^0 = \tfrac12 (\bfA\cdot\bfB - \bfC\cdot\bfE) 
\end{equation}
where $\bfE$, $\bfB$ are the electric and magnetic field vectors expressed in traditional spatial vector notation and $\bfA$, $\bfC$ are the spatial vectors with components $A_{\za}$, $C_{\za}$, and $J^0 = -J^a u_a$ is the helicity density. The spatial part of the helicity current is the spin density given by
\begin{equation}
   S_a \equiv J_{\za} = \tfrac12 (\bfE \times \bfA + \bfB \times \bfC)
                       +\tfrac12(C_0 \halfspace\bfE - A_0 \halfspace\bfB) \,,
\end{equation}
usually presented in transverse gauge \cite{Cameron_etal:2012}. Here $J_{\za} = h_a{}^b J_b$, where $h_{ab} = g_{ab} + u_a u_b$ is the spatial projection, cf. appendix \ref{appendix}. 
Written in terms of these quantities, the conservation law \eqref{Jcons} has the form
\begin{align}
   \bdot\Hscr + \pd_{\za} S^{\za} \cong 0
\end{align}
where here and in the following the overdot denotes time derivative, $\bdot\Hscr = u^a \pd_a \Hscr = \pd_0 \Hscr$.
This relation is the first level of the helicity hierarchy, a three level hierarchy of conservation laws
\begin{align}
 \bdot\Hscr + \pd_{\za} S^{\za} &\cong 0 \label{hh1} \\ 
 \bdot S_{\za} + \pd_{\quart\zb} \Sigma_{\za}{}^{\zb} &\cong 0 \label{hh2}\\
 \bdot\Sigma_{\za\zb} + \pd_{\zc} \halfspace N_{\za\zb}{}^{\zc} &\cong 0 \,
  \label{hh3}
\end{align}
where the spin flux (or infra-zilch \cite{Cameron_etal:2012}) is the symmetric spatial tensor
\begin{equation} \label{eq:Sigma}
   \Sigma_{ab} = h_{ab} \halfspace \Hscr 
                +C_{(\za} E_{\zb)} - A_{(\za} B_{\zb)}
\end{equation}
and the flux of infra-zilch is the spatial tensor
\begin{equation}
   N_{abc} = N_{(ab)c}
           = h_{ab} S_c + C_{(\za} \!\lrpd_{|\zc|} \halfspace A_{\zb)} 
           = h_{ab} S_c + C_{(\za} A_{\zb),\zc} - A_{(\za} C_{\zb),\zc}
\end{equation}
using the bidirectional-arrow notation.
Cameron et al.~ \cite{Cameron_etal:2012} defined a conserved rank-three object $\Nscr^{abc} =\Nscr^{(ab)c}$ that they called the helicity array having three spacetime indices. The array was given the components
\begin{align}
      \Nscr^{000} ={}& \Hscr \\[1pt]
      \Nscr^{0\za0} ={}& \Nscr^{00\za} = S^a \\[1pt]
      \Nscr^{\za\zb0} ={}& \Nscr^{0\za\zb} =  \Sigma^{ab} \\[1pt]
      \Nscr^{\za\zb\zc} ={}& N^{abc} \,,
\end{align}
leading to the conservation relation $\pd_c \Nscr^{abc} \cong0$. 
They also showed that the array, while not being a tensor, could be mapped to the zilch tensor \cite{Lipkin:1964,Morgan:1964,Kibble:1965} by the mapping
\begin{equation}\label{map}
   {\bf A} \rightarrow \nabla \times {\bf A} \,,\quad
   {\bf C} \rightarrow \nabla \times {\bf C}  \,.
\end{equation}
Despite this close relation to the zilch tensor, the helicity array has up to now not been reformulated in a Lorentz covariant form. In the next section we introduce a Lorentz covariant tensor that contains the same information as the helicity array.

\section{Helicity tensor in complex duality-symmetric formulation}\label{sec:Htensor}

Recall the formulation of Maxwell theory in terms of a complex vector potential $\Ascr_a = A_a + i\halfspace C_a$ with field strength $\Fscr_{ab} = -2\,\Ascr_{[a,b]}$, cf. \cite{Cameron&Barnett:2012,Bliokh_etal:2013,Aghapour_etal:2019} and references therein. 
We shall now define a Lorentz covariant complex tensor $\Hscr_{abc}$, which we call the helicity tensor, that contains the same information as the helicity array.  
Let 
\begin{align}\label{Ytensor}
\Yscr_{abc} = \tfrac{i}2\,\bar{\Ascr}_a \lrpd_{c} \,\Ascr_b \,. 
\end{align}
This tensor is manifestly duality invariant and is conserved with respect to the third index as we will show in the following. 

Taking the divergence of \eqref{Ytensor} we find
\begin{align}
\Yscr_{ab}{}^c{}_{,c} ={}& \tfrac{i}2 (\bar\Ascr_a\square \Ascr_b - \Ascr_b\square \bar{\Ascr}_a)  =  \tfrac{i}2 \bar{\Ascr}_a \Fscr^a{}_{b,a} - \tfrac{i}2 \Ascr_b \barFscr^b{}_{a,b} \nonumber \\ 
&\quad 
+ \tfrac{i}2 \bar{\Ascr}_a (\div \Ascr)_{,b} - \tfrac{i}2 \Ascr_b (\div \bar{\Ascr})_{,a}  \,,
\end{align}
where now $\div \Ascr$ is the spacetime divergence $\pd_a \Ascr^a$.
This shows that $\Yscr_{abc}$ is conserved in Lorenz gauge, \begin{equation}\label{eq:lorenz}
\div \Ascr = \div\bar{\Ascr}=0\,, 
\end{equation}
provided that the field equations $\Fscr^b{}_{a,b}=0$ and $\barFscr^b{}_{a,b}=0$ are satisfied.

Let 
\begin{align} 
P_{abcd} = g_{c(a}g_{b)d} -\tfrac{i}{2} \varepsilon_{abcd},
\end{align} 
where $g_{ab}$ is the Minkowski metric, and $\varepsilon_{abcd}$ is the totally antisymmetric Levi-Civita tensor, and define the operator $\hat P$ by its action on 2-tensors, $(\hat P\, t)_{ab} = P_{ab}{}^{cd}\, t_{cd}$. Due to $P_{abcd} P^{cdef} = g_{a}{}^{e} g_{b}{}^{f}$, we have that $(\hat P)^2$ is the identity operator, and hence $\hat P$ is an involution.  We now define the helicity tensor by 
\begin{align} 
\Hscr_{abc} = (\hat P \Yscr)_{abc} 
\end{align} 
where $\hat P$ acts on the first two indices of $\Yscr_{abc}$. Taking into account the definition of $\hat P$ we have that 
\begin{align} 
\pd_c \Hscr^{abc} \cong 0
\end{align} 
provided the vacuum Maxwell equations are satisfied and $\Ascr_a$ is in Lorenz gauge. As we shall see, $\Hscr_{abc}$ has odd parity. Since $\hat P$ is invertible, $\Hscr_{abc}$ contains the same information as $\Yscr_{abc}$. Therefore, in order to demonstrate that $\Hscr_{abc}$ contains the same information as $\Nscr_{abc}$, it is sufficient to consider $\Yscr_{abc}$.

The real part of $\Yscr_{abc}$ is symmetric in first two indices, while the imaginary part of it is antisymmetric in those indices. We call the real part $Y_{abc}$ and the imaginary part $X_{abc}$ so that
\begin{align}
    \Yscr_{abc} = Y_{abc} + i X_{abc}
\end{align}
We have 
\begin{align}\label{ReImparts}
Y_{abc}={}&\Re \{\Yscr_{abc}\} = \Yscr_{(ab)c} = \tfrac{i}2\,\bar{\Ascr}_{(a}\lrpd_{|c|}\,\Ascr_{b)} \, ,\\
X_{abc}={}&\Im \{\Yscr_{abc}\} = -i\,\Yscr_{[ab]c} = \tfrac12\,\bar{\Ascr}_{[a}\lrpd_{|c|}\,\Ascr_{b]}\ .
\end{align}
In terms of real potentials $A_a$ and $C_a$ we have
\begin{subequations} \label{eq:XY}
\begin{align}
   Y_{abc} &= C_{(a} \!\lrpd_{|c|} \halfspace A_{b)}
           = C_{(a} A_{b),c} - A_{(a} C_{b),c} \label{Yabc} \\
   X_{abc} &= \tfrac12(A_a \lrpd_c\, A_b + C_a \lrpd_c\, C_b) 
           = A_{[a} A_{b],c} + C_{[a} C_{b],c} \label{Xabc} \ .
\end{align}
\end{subequations} 
We now note that the helicity tensor has the form  
\begin{align} 
\Hscr_{abc} = Y_{abc} + \dual X_{abc}
\end{align} 
where $\dual X_{abc} = \tfrac{1}{2} \varepsilon_{ab}{}^{ef} X_{efc}$ is the dual of $X_{abc}$. It can be seen from \eqref{eq:XY} that both  $Y_{abc}$ and $\dual X_{abc}$ have odd parity.

We are now ready to prove that the helicity array and corresponding conservation laws are incorporated in the helicity tensor \eqref{Ytensor}. 
\begin{remark}\label{rem:gauge}
In previous work on the helicity array and related conservation laws, the transverse gauge \eqref{transverse_gauge} has been assumed. Here we use the Lorenz gauge \eqref{eq:lorenz} which is Lorentz invariant. On Minkowski spacetime, transverse gauge is consistent with Lorenz gauge, but breaks Lorentz invariance. 
\end{remark}

The helicity current $J^a$, which has a covariant form \eqref{Hcurrentvec}, is expected to have a covariant relationship with the helicity tensor. In fact, it can be written as
\begin{align}\label{Hcurrent-vs-Htensor}
    J^a ={}& \tfrac{i}4 (\bar\Fscr^{ab}\,\Ascr_b - \Fscr^{ab}\,\bar\Ascr_b) \nonumber\\
        ={}& -\tfrac12 \Yscr_b{}^{ba} +\tfrac{i}4 (\bar\Ascr^a\,\div \Ascr - \Ascr^a \,\div \bar\Ascr) + \tfrac{i}4( \bar\Ascr^b\,\Ascr^a - \Ascr^b\,\bar\Ascr^a)_{,b} \ ,
\end{align}
which shows that the helicity current $J^a$ is equivalent to the trace of the helicity tensor $\Hscr_b{}^{ba}=\Yscr_b{}^{ba}$, up to the Lorenz gauge and a trivial current (i.e.\ having a vanishing divergence). To be concrete, we introduce the equivalent helicity current
\begin{align}\label{eq:tildeJ_Htensor}
    \tilde J_a = \tfrac{i}4\, \bar\Ascr^b\,\lrpd_a\,\Ascr_{b} = -\tfrac12 \Yscr^b{}_{ba} = - \tfrac12 \Hscr^b{}_{ba} \ .
\end{align}
From \eqref{Hcurrent-vs-Htensor}, we find that
\begin{align}\label{Hcurrent-vs-Htensor2}
    J^a - \tilde J^a = \tfrac{i}4 (\bar\Ascr^a\,\div \Ascr - \Ascr^a \,\div \bar\Ascr) + \tfrac12\,K^a \ ,
\end{align}
where
\begin{align}\label{eq:K}
    K^a = i\,( \bar\Ascr^{[b}\,\Ascr^{a]})_{,b} = 2\,( C^{[b}\,A^{a]})_{,b}
\end{align}
has a vanishing divergence. Finally, we note that the trace of the helicity tensor $\Hscr_{abc}$ is given by the trace of $Y_{abc}$, because $\dual X_{abc}$ is antisymmetric in its first two indices.

Using \eqref{eq:tildeJ_Htensor} and performing the $1+3$ decomposition of $\tilde J_a$ (see appendix \ref{appendix2}) the helicity density in transverse gauge can be obtained as: 
\begin{align}
\Hscr \TG -\tfrac12\, Y_a{}^{a0} = -\tfrac12\, \Hscr_a{}^{a0}
\end{align}

The second level of the helicity array turns out to be related to $\dual X_{abc}$. In particular the spin density is related to the helicity tensor through the $1+3$ decomposition (see appendix \ref{appendix2}) by
\begin{align}
     S^{\za} \TG \dual X^{0\za0} = \Hscr^{0\za0} \ ,
\end{align}
while the second conservation law in \eqref{hh2} (i.e. the conservation of spin) is obtained from
\begin{align}
    \bdot S^{\zb} + \Sigma^{\zb c}{}_{,c} \TG \dual X^{0\zb c}{}_{,c} = \Hscr^{0\zb c}{}_{,c} 
    \cong 0  \,.
\end{align}
Finally, from \eqref{sigma-vs-Y} the infra-zilch can be constructed as 
\begin{align}
    \Sigma^{\za\zb} \TG h^{\za\zb}\,\Hscr  + Y^{\za\zb0} = -\tfrac12\,h^{\za\zb}\,\Hscr_{c}{}^{c 0}  + \Hscr^{\za\zb 0} \ ,
\end{align}
and its conservation law (i.e. the third level of conservation of the helicity array) is obtained from
\begin{align}
     \bdot\Sigma^{\za\zb} + N^{\za\zb\zc}{}_{,\zc} \TG - \tfrac12\,h^{\za\zb}\, \Hscr_c{}^{cd}{}_{,d} + \Hscr^{\za\zb c}{}_{,c} \cong 0 \ .
\end{align}
Finally, the flux of infra-zilch can be written as
\begin{align}
    N^{\za\zb\zc} \TG -\tfrac12\,h^{\za\zb}\, (\Hscr_d{}^{d\zc}-K^{\zc}) + \Hscr^{\za\zb\zc} \ ,
\end{align}
where $K^a$ is the null current \eqref{eq:K}. A summary of these relations between the helicity array and the helicity tensor is given in table \ref{table}.

\begin{table}[!t]
\begin{center}
\renewcommand{\arraystretch}{1.6}
\resizebox{\textwidth}{!}{
\begin{tabular}{lccc} 
\hline
& symbol & helicity array & helicity tensor \\ 
\hline
helicity density & $\Hscr$ & $\Nscr^{000}$ & $-\tfrac12\Hscr_a{}^{a0}$  \\
spin density & $S^{\za}$ & $\Nscr^{0\za0} = \Nscr^{00\za}$ & $-\Hscr^{0\za0} = -\tfrac12(\Hscr_b{}^{b\za}-K^{\za})$  \\
infra-zilch density & $\Sigma^{\za\zb}$ & $\Nscr^{\za\zb 0} = \Nscr^{0\za\zb}$ &  $ -\tfrac12\,h^{\za\zb}\,\Hscr_{c}{}^{c 0}  + \Hscr^{\za\zb 0} = -(\Hscr^{0\za\zb} - \tilde K^{\za\zb})$  \\
infra-zilch flux & $N^{\za\zb\zc}$ & $\Nscr^{\za\zb\zc}$ & $-\tfrac12\,h^{\za\zb}\, (\Hscr_d{}^{d\zc}-K^{\zc}) + \Hscr^{\za\zb\zc}$
\\[2pt]
\hline
\end{tabular}
}
\\[.3cm]

\caption{\small{Summary of the relations between the helicity tensor and the helicity array. Transverse gauge is assumed. In the second and third rows the left hand side is the charge density while the right hand side is the flux of the previous charge. This fact illustrates the link between the conservation laws. The minus signs in the last column are due to the definition of the helicity tensor with indices down.}}
\label{table}
\end{center}

\end{table}
\section{Noether analysis}\label{sec:Noether}
Using a complex vector potential $\Ascr_a = A_a + i\halfspace C_a$, Maxwell's equations can be derived from the duality-symmetric Lagrangian (\emph{cf.~}\cite{Bliokh_etal:2013})
\begin{equation}\label{ds_lag}
   \Lscr = -\tfrac18 \Fscr_{ab} \barFscr^{ab} 
         = -\tfrac12\kappa^{abcd} \Ascr_{a,b} \, \barAscr_{c,d}
\end{equation}
where $\Fscr_{ab} = -2\,\Ascr_{[a,b]}$ and the bar denotes complex conjugation.
Here, $\kappa_{abcd}= g_{c[a} g_{b]d}$ is the antisymmetry projector.\footnote{This tensor projects out the antisymmetric part of any index pair, e.g.\ $M_{[ab]} = \kappa_{ab}{}^{cd} M_{cd}$. It has the same index symmetries as the Riemann tensor, $\kappa_{abcd} = \kappa_{cdab} = \kappa_{[ab]cd}$ and $\kappa_{a[bcd]} =0$.}

In \cite{Bliokh_etal:2013}, the real and imaginary parts of $\Ascr_a$ were taken as the basic field variables. Here, we will instead use $\Ascr_a$ and $\bar\Ascr_a$ as the basic fields. Since they represent linearly independent combinations of $A_a$ and $C_a$, they can be used as alternative field variables. They carry the same number of degrees of freedom (4 each) and are treated as formally independent. 
Defining $\Mscr^a = \Esf^a_\Ascr(\Lscr)$,  where $\Esf_\Xscr$ is the Euler operator for a variable $\Xscr$, the Euler-Lagrange expressions are given by
\begin{equation}\label{EL}
 \begin{split}
   \Mscr^a &= -\frac{\pd}{\pd x^b} \frac{\pd\Lscr}{\pd\Ascr_{a,b}}
            = - \tfrac14\barFscr^{ab}{}_{,b}
 \end{split}                  
\end{equation}
and $\Esf_{\!\bar\Ascr}^a(\Lscr) = \bar \Mscr^a$.
Expressed in terms of the potentials they become
\begin{equation}
    \Mscr_a = \tfrac14 \square \barAscr_a -\tfrac14 (\div\barAscr)_{,a} \,.
\end{equation}
Here $\div \Ascr = \partial_a \Ascr^a$ is the divergence. 

\begin{align}\label{Ydiv}
\Yscr_{ab}{}^c{}_{,c} ={}& \tfrac{i}2 (\bar\Ascr_a\square \Ascr_b - \Ascr_b\square \bar{\Ascr}_a)  =  2 i \bar{\Ascr}_a \bar{\Mscr}_b -2 i \Ascr_b \Mscr_a \nonumber \\
&\quad + \tfrac{i}2 \bar{\Ascr}_a (\div \Ascr)_{,b} - \tfrac{i}2 \Ascr_b (\div \bar{\Ascr})_{,a}  \,.
\end{align}

The presence of the gauge dependent terms, the equation \eqref{Ydiv} does not as it stands lead to a conservation law in the characteristic form needed for correspondence with Noether theory (see \cite{Aghapour_etal:2019,Olver:1993} for details of our use of the Noether formalism). One way to find such a form is to modify the Lagrangian by adding terms which take care of the gauge constraints. 
To this end we define a modified Lagrangian depending on an auxiliary complex dependent variable $\omega$ 
\begin{equation}\label{modifiedL}
   \tilde\Lscr = \Lscr + \omega \div \Ascr + \bar\omega \div \bar\Ascr 
   \,.
\end{equation}
Since the additional parts are linear in $\Ascr_{a,b}$ and $\bar\Ascr_{a,b}$, no new terms enter in the Euler-Lagrange expressions \eqref{EL}.
The modified Lagrangian \eqref{modifiedL} leads to the complex Euler-Lagrange expression
\begin{equation}
   \Omega =  \Esf_\omega(\tilde\Lscr)
          = \frac{\pd \tilde\Lscr}{\pd\omega} = \div \Ascr \\[3pt]
\end{equation}
and its complex conjugate in addition to those in \eqref{EL}.
This shows that the Lagrangian $\tilde\Lscr$ gives the duality-symmetric Maxwell equations in Lorenz gauge.
The relation \eqref{Ydiv} can then be written in the form
\begin{equation}\label{Ydiv2}
        \Yscr_{ab}{}^c{}_{,c} =   2i\, \bar{\Ascr}_a \,\bar{\Mscr}_b -2 i\, \Ascr_b\, \Mscr_a + \tfrac{i}2\,\bar{\Ascr}_a\, \Omega_{,b} - \tfrac{i}2\,\Ascr_b\, \bar\Omega_{,a} \,.
\end{equation} 
As in the case of the zilch conservation law (see \cite{Aghapour_etal:2019}), there are Euler-Lagrange expressions ($\Omega$ and $\bar\Omega$) that are differentiated in \eqref{Ydiv2}. Therefore to arrive at a conservation law in characteristic form we need to first perform partial integrations leading to
\begin{align}
    \Yscr_{ab}{}^c{}_{,c} ={}& 2 i\, \bar{\Ascr}_a \,\bar{\Mscr}_b -2 i\, \Ascr_b\, \Mscr_a - \tfrac{i}2\, \bar{\Ascr}_{a,b}\, \Omega \nonumber \\ 
&\quad + \tfrac{i}2\, \Ascr_{b,a}\, \bar\Omega + \tfrac{i}2\, (\bar{\Ascr}_a\, \Omega)_{,b} - \tfrac{i}2\, (\Ascr_b\, \bar\Omega)_{,a} .                
\end{align}
Defining a tensor which is equivalent\footnote{In the sense of being equal for solutions of the $\tilde\Lscr$ field equations.} to $\Yscr_{abc}$ by
\begin{equation} \label{eq:Ytilde}
   \tilde \Yscr_{abc} = \Yscr_{abc} - \tfrac{i}2\,g_{bc}\, \bar{\Ascr}_a \,\div \Ascr  + \tfrac{i}2\, g_{ac} \,\Ascr_b\, \div \bar\Ascr 
\end{equation}
the conservation law can be written in the characteristic form
\begin{equation}\label{charform}
   \tilde \Yscr_{ab}{}^c{}_{,c} = 2 i\, \bar{\Ascr}_a \,\bar{\Mscr}_b -2 i\, \Ascr_b\, \Mscr_a - \tfrac{i}2\, \bar{\Ascr}_{a,b}\, \Omega + \tfrac{i}2\, \Ascr_{b,a}\, \bar\Omega \,.
\end{equation}

From equation \eqref{charform} we can read off the characteristic functions, which are the coefficients of the Euler-Lagrange expressions:
\begin{align}
Q^{\ms\Mscr}_{abc} = - 2\, i\, \Ascr_b\,g_{ac} \ , \quad Q^{\ms{\bar{\Mscr}}}_{abc} = 2\, i\, \bar\Ascr_a\,g_{bc} \ , \quad Q^{\ms \Omega} = - \tfrac{i}2\, \bar{\Ascr}_{a,b} \ , \quad Q^{\ms{\bar\Omega}} = \tfrac{i}2\, \Ascr_{b,a}
\end{align}
The corresponding symmetry generator then becomes
\begin{align}\label{symgen}
   v_{ab} ={}& Q^{\ms\Mscr}_{abc} \frac{\pd}{\pd \Ascr_c} + Q^{\ms{\bar{\Mscr}}}_{abc} \frac{\pd}{\pd \barAscr_c}  + Q^{\ms \Omega}_{ab} \frac{\pd}{\pd \omega} + Q^{\ms{\bar\Omega}}_{ab} \frac{\pd}{\pd\bar\omega} \nonumber\\
   ={}& - 2\, i\, \Ascr_b\,g_{ac}\, \frac{\pd}{\pd \Ascr_c}
           +2\, i\, \bar\Ascr_a\,g_{bc}\, \frac{\pd}{\pd \bar\Ascr_c} - \tfrac{i}2\, \bar{\Ascr}_{a,b} \frac{\pd}{\pd \omega} + \tfrac{i}2\, \Ascr_{b,a} \frac{\pd}{\pd\bar\omega}
          \,.
\end{align}
The first prolongation of the generator \eqref{symgen} is given by
\begin{align}
   \pr v_{ab} ={}& v_{ab} + Q^{\ms\Mscr}_{abc,d} \,\frac{\pd}{\pd \Ascr_{c,d}} + Q^{\ms{\bar{\Mscr}}}_{abc,d}\, \frac{\pd}{\pd \barAscr_{c,d}} + Q^{\ms \Omega}_{ab,c} \frac{\pd}{\pd \omega_{,c}} + Q^{\ms{\bar\Omega}}_{ab,c} \frac{\pd}{\pd\bar\omega_{,c}} \nonumber\\
   ={}& v_{ab} - 2\, i\, \Ascr_{b,d}\,g_{ac}\, \frac{\pd}{\pd \Ascr_{c,d}}
                       +2\, i\, \bar\Ascr_{a,d}\,g_{bc}\, \frac{\pd}{\pd \bar\Ascr_{c,d}} \nonumber \\
    &\quad - \tfrac{i}2\, \bar{\Ascr}_{a,bc} \frac{\pd}{\pd \omega_{,c}} + \tfrac{i}2\, \Ascr_{b,ac} \frac{\pd}{\pd\bar\omega_{,c}}\,.
\end{align}
The action of the prolonged generator on the duality-symmetric Lagrangian \eqref{ds_lag} is a total divergence as
\begin{align}
   \pr v_{ab}(\Lscr) ={}& Q^{\ms\Mscr}_{abc,d} \,\frac{\pd \Lscr}{\pd \Ascr_{c,d}} + Q^{\ms{\bar{\Mscr}}}_{abc,d}\, \frac{\pd \Lscr}{\pd \barAscr_{c,d}} + Q^{\ms \Omega}_{ab} \frac{\pd\Lscr}{\pd \omega} + Q^{\ms{\bar\Omega}}_{ab} \frac{\pd\Lscr}{\pd\bar\omega}\nonumber\\
   ={}& - 2\, i\, \Ascr_{b,d}\,g_{ac}\, (\tfrac14 \Fscr^{cd})
                     +2\, i\, \bar\Ascr_{a,d}\,g_{bc}\, (\tfrac14 \bar\Fscr^{cd}) \nonumber \\
        &\quad - \tfrac{i}2\, \bar{\Ascr}_{a,b} (\div\Ascr) + \tfrac{i}2\, \Ascr_{b,a} (\div\bar\Ascr) \nonumber\\
    ={}& \Uscr_{ab}{}^c{}_{,c}
\end{align}
where
\begin{equation}
   \Uscr_{abc} = \tfrac{i}2\,\bar\Ascr_a\,\Ascr_{c,b} - \tfrac{i}2 \Ascr_b\bar\Ascr_{c,a}-\tfrac{i}2\, g_{bc}\,\bar\Ascr_a\,\div\Ascr + \tfrac{i}2\, g_{ac}\,\Ascr_b\,\div\bar\Ascr  \,.
\end{equation}
This leads to the conserved tensor
\begin{align}\label{Htensorsymm}
  -Q^{\ms\Mscr}_{abd}\,\frac{\pd \Lscr}{\pd \Ascr_{d}{}^{,c}} - Q^{\ms{\bar{\Mscr}}}_{abd}\, \frac{\pd \Lscr}{\pd \bar\Ascr_{d}{}^{,c}} + \Uscr_{abc} 
 ={}& \Yscr_{abc} - \tfrac{i}2\,g_{bc}\, \bar\Ascr_a \,\div \Ascr  + \tfrac{i}2\, g_{ac} \,\Ascr_b\, \div \bar\Ascr \nonumber  \\ 
 ={}& \tilde \Yscr_{abc} \ ,
\end{align}
as given in \eqref{eq:Ytilde}. 
We note that $\Yscr_{abc}$ and $\tilde \Yscr_{abc}$ are equal modulo Lorenz gauge as expected. In the present version of the Noether formalism, this corresponds to the vanishing of their difference  on-shell. Such a quantity is itself considered as a trivial conservation law of the first kind (see \cite{Olver:1993}). The conservation laws of $\Yscr_{abc}$ and $\tilde \Yscr_{abc}$ are then said to be equivalent.

The symmetries underlying the helicity hierarchy of conserved quantities, including helicity, spin, and infra-zilch, have been discussed by Cameron et al. \cite{Cameron2012,Cameron_etal:2012} and Bliokh et al. \cite{Bliokh_2014}, in transverse gauge. The analysis above gives for the first time a derivation of the helicity hierarchy of conserved currents as Noether currents arising from a variational symmetry of a duality symmetric action. 

It is instructive to notice the symmetric and antisymmetric parts of the generator \eqref{symgen}, which are, respectively, responsible for the conservation of real and imaginary parts of the helicity tensor in \eqref{ReImparts}. The symmetric part is
\begin{align}\label{eq:symgen}
   v_{(ab)} ={}&  - 2\, i\, \Ascr_{(a}\,g_{b)c}\, \frac{\pd}{\pd \Ascr_c} - \tfrac{i}2\, \bar{\Ascr}_{(a,b)}\frac{\pd}{\pd \omega} + c.c. \nonumber\\
          ={}& 2\,C_{(a}\,g_{b)c}\, \frac{\pd}{\pd A_c} -2\,A_{(a}\,g_{b)c}\, \frac{\pd}{\pd C_c} + \tfrac12\,C_{(a,b)}\frac{\pd}{\pd \omega_{\ms R}} -\tfrac12\,A_{(a,b)}\frac{\pd}{\pd \omega_{\ms I}}
          \,.
\end{align}
while the antisymmetric part is
\begin{align}\label{eq:symgen2}
   v_{[ab]} ={}& 2\, i\, \Ascr_{[a}\,g_{b]c}\, \frac{\pd}{\pd \Ascr_c} - \tfrac{i}2\, \bar{\Ascr}_{[a,b]}\frac{\pd}{\pd \omega} - c.c.
   \nonumber\\
          ={}& i \Big[2\,A_{[a}\,g_{b]c}\, \frac{\pd}{\pd A_c} +2\,C_{[a}\,g_{b]c}\, \frac{\pd}{\pd C_c} - \tfrac12\,A_{[a,b]}\frac{\pd}{\pd \omega_{\ms R}} -\tfrac12\,C_{[a,b]}\frac{\pd}{\pd \omega_{\ms I}} \Big]
          \,.
\end{align}
The symmetric part \eqref{eq:symgen} is associated to the conservation of helicity and infra-zilch, whereas the antisymmetric part underlies the conservation of the spin part of the angular momentum. The equation \eqref{eq:symgen2} gives the  symmetry underlying the conservation of the spin part of the angular momentum of the electromagnetic field. This completes the analysis of the helicity hierarchy of conservation laws as Noether currents, cf.  earlier discussions in \cite{Cameron_etal:2012,Cameron2012,Bliokh_2014}.

The duality transformation underlying the helicity conservation law \eqref{Jcons} is included in the symmetry transformation generated by \eqref{symgen}. It can be realized by taking the trace of that generator resulting in
\begin{align}
    v_{\scriptscriptstyle{\mathcal H}} &= - 2\, i\, \Ascr_a\, \frac{\pd}{\pd \Ascr_a}
           +2\, i\, \bar\Ascr_a\, \frac{\pd}{\pd \bar\Ascr_a} \nonumber\\
        &= 2\,C_a\, \frac{\pd}{\pd A_a}  - 2\,A_a\, \frac{\pd}{\pd C_a} \ ,
\end{align}
which generates the duality transformation.


\section{Concluding Remarks}

We have introduced a new Lorentz covariant tensor, $\Hscr_{abc}$, which we call the helicity tensor, that is conserved in Lorenz gauge, and contains the same 
information as the helicity array of Cameron et al. \cite{Cameron_etal:2012}. In particular, the 
conserved currents expressing helicity, spin, and infra-zilch can be obtained from the helicity tensor 
by performing a $1+3$ decomposition and specializing to transverse gauge. The fact that helicity, spin and infra-zilch are observer dependent parts of a Lorentz covariant object, the helicity tensor, is analogous to the fact that spin and orbital angular momentum are parts of the Lorentz covariant total angular momentum. With this in mind, it would be interesting to investigate if there are observer dependence properties of these quantities which are analogous to the spin-orbit interaction discussed by Bliokh et al. \cite{2015NaPho...9..796B} and Smirnova et al. \cite{PhysRevA.97.043840}.  

The construction of the helicity tensor is carried out in terms of the duality-symmetric formulation of Maxwell theory and is local with respect to the pair of electromagnetic potentials $A_a, C_a$. However, as in any other treatment of the conservation laws expressed in the helicity array, the construction is non-local with respect to the standard formulation in terms of a single electromagnetic potential. Indeed, the intrinsic non-locality of these conserved quantities is emphasized by the work of  I.\ and Z.\  Bialynicki-Birula \cite{Bialynicki-Birula2011} who showed that the spin and orbital parts of the angular momentum of the electromagnetic field are given in a gauge invariant manner in terms of vector potentials constructed from the Maxwell field strength using a non-local transformation. It appears worthwhile to consider whether a similar construction can be carried out for the other conserved quantities contained in the helicity tensor.

We have further demonstrated that the helicity tensor and its associated conservation laws correspond to Noether currents associated to  variational symmetries of the duality-symmetric Lagrangian for Maxwell theory, amended with  gauge fixing terms. Symmetries giving rise to the helicity array components as Noether quantities were discussed in papers by Cameron et al. \cite{Cameron2012,Cameron_etal:2012} and Bliokh et al. \cite{Bliokh_2014}. The analysis in the just cited papers was restricted to the transverse gauge, and the variational aspect of the symmetries were not investigated. The symmetry generator given here is new, and differs from the non-local symmetries discussed in these papers. It would be interesting to understand the relation between these different symmetries. In particular, the physical interpretation of the symmetries presented here should be clarified.

Here we have defined the the helicity tensor $\Hscr_{abc}$ by applying the parity modifier operator $\hat P$ to the complex conserved tensor $\Yscr_{abc}$, obtaining a real tensor with the desired parity properties. It would be interesting to discuss the complex nature of  $\Yscr_{abc}$ and its parity properties in the framework of geometrical algebra of the 4-dimensional spacetime, cf. \cite{DRESSEL20151}. We plan to return to this subject in a future publication.

Finally, we mention that it is of interest to investigate whether some of the results considered here can be generalized to Maxwell theory in more general spacetimes, in particular to analyze the generalization of the helicity hierarchy of conservation laws to certain algebraically special spacetimes. The fact that the helicity tensor is conserved in Lorenz gauge may allow us to apply the construction to interesting families of solutions of Maxwell theory, including for example Hopfions and several types of optical beams, cf.\ eg. \cite{2008NatPh...4..817I,2013PhRvL.111o0404K,2017PhR...667....1A,2018CQGra..35x5010S} and references therein. 

\appendix
\section{A covariant notation for $1+3$ decomposition}\label{appendix}
Performing a time-space decomposition in any relativistic theory is crucial when considering experimental and observational measurements. It is necessary since measurements depend on the rest space of a lab or an observer (or equivalently on the 4-velocity of the rest space of the lab or observer). This is perhaps most often done by introducing 3-dimensional indices which only take values which number the coordinate axes in the observer's rest space. While perfectly valid, this procedure does not respect the spacetime covariance of the underlying relativistic theory. Also, it obscures the fact that measurements depend on the world line of a lab or an observer. In this note we will instead use a slightly different approach which is fully spacetime covariant but in which the time-space decomposition can
nevertheless be just as clearly displayed. This approach also has the advantage of employing only a single type of indices.

The covariant decomposition will be done by using projection tensors which can project any tensor into either the worldline of the observer or into the observer's rest space. Starting with projection into the rest space, the projection tensor is defined by
\begin{equation}\label{hab}
   h^a{}_b = g^a{}_{b} + u^a u_b
\end{equation}
where $u^a$ is the 4-velocity of the observer and $g^a{}_{b}$ is the 4-dimensional Kronecker delta. The latter is usually written with a 
``$\delta$" but we avoid this notation here since it can be in conflict with $h^a{}_b$ which can serve as a Kronecker delta in the rest space. The projection into the observer's worldline has the form
\begin{equation}\label{fab}
   f^a{}_b = -u^a u_b \,.
\end{equation}
The projection property of the tensors defined in \eqref{hab} and \eqref{fab} is manifested by the relations
\begin{equation}
   h^a{}_c h^c{}_b = h^a{}_b \,,  \qquad  f^a{}_c f^c{}_b = f^a{}_b \,.
\end{equation}
Taking a general vector $V^a$ as an example, its decomposition in time and space parts takes the form
\begin{equation}\label{decomp}
   V^a = f^a{}_b V^b + h^a{}_b V^b 
       = -u_b V^b u^a + V^{\za}
\end{equation}
where the barred index is used to denote a spatially projected index
\begin{equation}
   V^{\za} = h^a{}_b V^b \,.
\end{equation}
Note that using this notation, intrinsically spatial tensors such as $h_{ab}$ or the electric and magnetic fields for example can be written both with and without bars, for example $E_{\za} = E_a$.
It also follows from its definition that a barred index can be raised and lowered with respect to either $g_{ab}$ or $h_{ab}$, for example
\begin{equation}
   V^{\za} = g^{ab} V_{\zb} = h^{ab} V_{\zb} \,.
\end{equation}
Employing an observer adapted reference frame, a projection of a vector $V^a$ into the worldline of the observer has the single component $V^0$, which from the relation \eqref{decomp} can be identified as the scalar $V^0 = -u_a V^a$. In the same way, lowering the free index in \eqref{decomp} gives $V_0 = -V^0 = u_a V^a$. 
We finally note that using this formalism allows spatial objects to be represented in a fully spacetime covariant way. Although this has been demonstrated here for vectors, it is clear that the formalism can be straightforwardly extended to any tensorial objects.

\section{The $1+3$ decomposition of the helicity tensor}\label{appendix2}
We start with the $1+3$ decomposition of the trace of the helicity tensor. Performing the $1+3$ decomposition of $\tilde J_a$ gives the following components which are identical to the helicity vector components in transverse gauge: 
\begin{align}
\tilde J_0 ={}& -\tfrac12 (C^a\, \dot A_a - A^a\, \dot C_a) \TG{} \tfrac12 ( \bfC\cdot\bfE - \bfA\cdot\bfB) = - \Hscr \\
\tilde J_{\za} ={}& -\tfrac12 ( C^b \, A_{b,\za} - A^b\, C_{b,\za}) \TG{} \tfrac12 (\bfB \times \bfC + \bfE \times \bfA) - \tfrac12 (C^{\zb}\, A_{\za,\zb} - A^{\zb}\, C_{\za,\zb} ) 
\end{align}

As the next task, we perform the $1+3$ decomposition of the real part $Y_{abc}$ of the helicity tensor. This gives
\begin{align}
        Y_{000} &=  C_0 \halfspace \bdot A_0 
                  - A_0 \halfspace \bdot C_0 \TG 0 \\[1pt]
      Y_{00\zc} &=  C_0 (\nabla A_0)_{\zc} 
                      - A_0 (\nabla C_0)_{\zc} \TG 0 \\[1pt]
      Y_{\za00} &= \tfrac12( C_0 \bdot A_{\za} - \bdot C_0 A_{\za} 
                  - A_0 \,\bdot C_{\za} + \bdot A_0 \,C_{\za}) \TG 0 \\[1pt]
    Y_{\za\zb0} &= C_{(\za} \bdot A_{\zb)} - A_{(\za} \bdot C_{\zb)}
               \TG -C_{(\za} E_{\zb)} + A_{(\za} B_{\zb)}
                 = h_{ab} \Hscr - \Sigma_{ab} \label{sigma-vs-Y}\\[1pt]
    Y_{0\zb\zc} &=  \tfrac12 ( C_0 A_{\zb,\zc} - A_{\zb}\halfspace  C_{0,\zc}  
                    - A_0\, C_{\zb,\zc} + C_{\zb} \, A_{0,\zc}) \TG 0 \\[1pt]
  Y_{\za\zb\zc} &= C_{(\za} A_{\zb),\zc} - A_{(\za} C_{\zb),\zc}  \,.
\end{align}
We see that only the parts $Y_{\za\zb0}$ and $Y_{\za\zb\zc}$, with 6 and 18 components respectively, are non-zero in transverse gauge. 

%
Looking at the 1+3 decomposition of the divergences of $Y_{abc}$, the only nontrivial part is given by
\begin{align}
   Y_{\za\zb}{}^c{}_{,c} 
      &= C_{(\za} \square A_{\zb)} - A_{(\za}
         \square C_{\zb)} \cong 0 \,,
\end{align}
where the weak equality ``$\,\cong\,$" refers to equality modulo the Euler-Lagrange equations of $\LRm$.

In the end, performing the 1+3 decomposition of $X_{abc}$ gives the following parts which are not identically zero
\begin{align}
        X_{\za00} &= A_{[\za} \bdot A_{0]} + C_{[\za} \bdot C_{0]} \TG0\\[2pt]
      X_{\za\zb0} &= \tfrac12(A_{\za} \bdot A_{\zb} - \bdot A_{\za} A_{\zb} + C_{\za} \bdot C_{\zb} - \bdot C_{\za} C_{\zb}) 
                 \TG E_{[a} A_{\zb]} + B_{[a} C_{\zb]} \\
      X_{0\zb\zc} &= A_{[0} A_{\zb],\zc} + C_{[0} C_{\zb],\zc} \TG 0 \\[2pt]
    X_{\za\zb\zc} &= A_{[\za} A_{\zb],\zc} + C_{[\za} C_{\zb],\zc}
\end{align}
For $\dual X_{abc}$, the 1+3 parts which are not identically zero are\footnote{\ Here $\eps_{\za\zb\zc}$ is the spatial Levi-Civita tensor.}
\begin{align}
     \dual X_{\za00} &= \tfrac12 \eps_{\za}{}^{\zb\zc} X_{\zb\zc0} 
     = \tfrac14 \eps_{\za}{}^{\zb\zc} (A_{\zb} \bdot A_{\zc} - \bdot A_{\zb} A_{\zc} + C_{\zb} \bdot C_{\zc} - \bdot C_{\zb} C_{\zc}) \nonumber \\
     &\TG \tfrac12\eps_{\za}{}^{\zb\zc} (E_{\zb} A_{\zc} + B_{\zb} C_{\zc})
     = S_a \\[1pt]
   \dual X_{\za\zb0} &= \tfrac12 \eps_{\za\zb}{}^{cd} X_{cd0}
                      = \eps_{\za\zb}{}^{\zc} X_{\zc00}  = \eps_{\za\zb}{}^{\zc} (A_{[\zc} \bdot A_{0]} + C_{[\zc} \bdot C_{0]} ) \TG 0 \\[1pt]
   \dual X_{0\zb\zc} &= \tfrac12 \eps_{0\zb}{}^{\zd\ze} X_{\zd\ze\zc} =  -\tfrac12\eps_{\zb}{}^{\zd\ze}
                         ( A_{\zd} \,A_{\ze,\zc}
                          +C_{\zd} \,C_{\ze,\zc}) \label{eq:*X0ab} \\[1pt]
 \dual X_{\za\zb\zc} &= \tfrac12 \eps_{\za\zb}{}^{de} X_{de\zc}
                      = \eps_{\za\zb}{}^{\ze} X_{0\ze\zc} = \eps_{\za\zb}{}^{\ze} (A_{[0} A_{\ze],\zc} + C_{[0} C_{\ze],\zc}) \TG 0 \,.
\end{align}

The antisymmetric part of $\dual X_{0\za\zb}$ ca be obtained as
\begin{align}\label{eq:X0ab_ASpart}
    \dual X_{0[\za\zb]} ={} & \tfrac14\,\epsilon_{\za\zb}{}^{\zc}\,(A^{\zd}\,A_{\zc,\zd}+C^{\zd}\,C_{\zc,\zd})  -\tfrac14\,\epsilon_{\za\zb\zd}\,(A^{\zd}\,A_{\zc}{}^{,\zc} + C^{\zd}\,C_{\zc}{}^{,\zc})
\end{align}
Based on \eqref{eq:X0ab_ASpart} and up to the Lorenz gauge, one can show that \eqref{eq:*X0ab} can be written in the form
\begin{align}\label{eq:X0ab_Sigma}
    \dual X_{0\za\zb} = \Sigma_{\za\zb} - \tilde K_{\za\zb} \ ,
\end{align}
where $\Sigma_{\za\zb}$ is the infra-zilch, Eq.\ \eqref{eq:Sigma} and
\begin{align}
    \tilde K_{\za\zb} =\tfrac14\left(2\,\epsilon_{\za[\zb}{}^{\zd}\,A_{\zc]}\,A_{\zd} + 2\,\epsilon_{\za[\zb}{}^{\zd}\,C_{\zc]}\,C_{\zd} - \epsilon_{\zb\zc}{}^{\zd}(A_{\zd}\,A_{\zd} + C_{\zd}\,C_{\za})\right)^{,\zc}\,,
\end{align}
which obviously has vanishing divergence with respect to the index $\zb$. 

Taking the divergence of $\dual X_{abc}$, there are two parts not identically zero given by
\begin{align}
    \dual X^{0\zb c}{}_{,c}
     &= \dual X^{0\zb0}{}_{,0} +\dual X^{0\zb\zc}{}_{,\zc} \nonumber \\
      &\TG -(\bdot S^b + \Sigma^{bc}{}_{,c}) \,. \\[1pt]
   \dual X^{\za\zb c}{}_{,c} 
     &= \dual X^{\za\zb0}{}_{,0} + \dual X^{\za\zb\zc}{}_{,\zc} \TG 0 \,.
\end{align}


\newcommand{\mnras}{Monthly Notices of the Royal Astronomical Society }

\newcommand{\arxivref}[1]{\href{http://www.arxiv.org/abs/#1}{{arXiv.org:#1}}}
\newcommand{\prd}{Phys. Rev. D} 
\newcommand{\pra}{Phys. Rev. A}

\bibliographystyle{unsrt}
\bibliography{references}

\begin{thebibliography}{10}

\bibitem{Cameron_etal:2012}
R.~P. Cameron, S.~M. Barnett, and A.~M. Yao.
\newblock {Optical helicity, optical spin and related quantities in
  electromagnetic theory}.
\newblock {\em New Journal of Physics}, 14, 2012.

\bibitem{1992PhRvA..45.8185A}
L.~{Allen}, M.~W. {Beijersbergen}, R.~J.~C. {Spreeuw}, and J.~P. {Woerdman}.
\newblock {Orbital angular momentum of light and the transformation of
  Laguerre-Gaussian laser modes}.
\newblock {\em Physical Review A}, 45:8185--8189, June 1992.

\bibitem{1994JMOp...41..963V}
S.~J. {van Enk} and G.~{Nienhuis}.
\newblock {Commutation Rules and Eigenvalues of Spin and Orbital Angular
  Momentum of Radiation Fields}.
\newblock {\em Journal of Modern Optics}, 41:963--977, May 1994.

\bibitem{Bialynicki-Birula2011}
I.~Bialynicki-Birula and Z.~Bialynicka-Birula.
\newblock {Canonical separation of angular momentum of light into its orbital
  and spin parts}.
\newblock {\em Journal of Optics}, 13(6):064014, 2011.

\bibitem{Bliokh_2014}
Konstantin~Y Bliokh, Justin Dressel, and Franco Nori.
\newblock Conservation of the spin and orbital angular momenta in
  electromagnetism.
\newblock {\em New Journal of Physics}, 16(9):093037, sep 2014.

\bibitem{Barnett_2016}
Stephen~M Barnett, L~Allen, Robert~P Cameron, Claire~R Gilson, Miles~J Padgett,
  Fiona~C Speirits, and Alison~M Yao.
\newblock On the natures of the spin and orbital parts of optical angular
  momentum.
\newblock {\em Journal of Optics}, 18(6):064004, apr 2016.

\bibitem{Bliokh2015}
K.~Y. Bliokh, F.~J. Rodríguez-Fortuño, F.~Nori, and A.~V. Zayats.
\newblock {Spin-orbit interactions of light}.
\newblock {\em Nature Photonics}, 9:796–808, December 2015.

\bibitem{Trueba_1996}
Jos{\'{e}}~L Trueba and Antonio~F Ra{\~{n}}ada.
\newblock The electromagnetic helicity.
\newblock {\em European Journal of Physics}, 17(3):141--144, may 1996.

\bibitem{Deser1976}
Stanley Deser and Claudio Teitelboim.
\newblock Duality transformations of abelian and non-abelian gauge fields.
\newblock {\em Phys. Rev. D}, 13:1592--1597, Mar 1976.

\bibitem{Jungho2020}
Jungho Mun, Minkyung Kim, Younghwan Yang, Trevon Badloe, Jincheng Ni, Yang
  Chen, Cheng-Wei Qiu, and Junsuk Rho.
\newblock Electromagnetic chirality: from fundamentals to nontraditional
  chiroptical phenomena.
\newblock {\em Light: Science \& Applications}, 9:2047--7538, 2020.

\bibitem{PhysRevB.99.174310}
Konstantin~Y. Bliokh and Franco Nori.
\newblock Spin and orbital angular momenta of acoustic beams.
\newblock {\em Phys. Rev. B}, 99:174310, May 2019.

\bibitem{Burns_2020}
Lucas Burns, Konstantin~Y Bliokh, Franco Nori, and Justin Dressel.
\newblock Acoustic versus electromagnetic field theory: scalar, vector, spinor
  representations and the emergence of acoustic spin.
\newblock {\em New Journal of Physics}, 22(5):053050, jun 2020.

\bibitem{Barnett2014}
S.~M. Barnett.
\newblock {Maxwellian theory of gravitational waves and their mechanical
  properties}.
\newblock {\em New Journal of Physics}, 16, 2014.

\bibitem{Andersson_etal:2018}
Sajad {Aghapour}, Lars {Andersson}, and Reebhu {Bhattacharyya}.
\newblock {Helicity and spin conservation in Maxwell theory and Linearized
  Gravity}.
\newblock {\em arXiv e-prints}, page arXiv:1812.03292, December 2018.

\bibitem{Lipkin:1964}
D.~M. Lipkin.
\newblock {Existence of a New Conservation Law in Electromagnetic Theory}.
\newblock {\em Journal of Mathematical Physics}, 5(5):696--700, 1964.

\bibitem{Morgan:1964}
Thomas~A. {Morgan}.
\newblock {Two Classes of New Conservation Laws for the Electromagnetic Field
  and for Other Massless Fields}.
\newblock {\em Journal of Mathematical Physics}, 5(11):1659--1660, November
  1964.

\bibitem{Kibble:1965}
Kibble T.~W. B.
\newblock {Conservation Laws for Free Fields}.
\newblock {\em Journal of Mathematical Physics}, 6(7):1022--1026, 1965.

\bibitem{MR1885280}
Stephen~C. Anco and Juha Pohjanpelto.
\newblock Classification of local conservation laws of {M}axwell's equations.
\newblock {\em Acta Appl. Math.}, 69(3):285--327, 2001.

\bibitem{Tang2010}
Yiqiao Tang and Adam~E. Cohen.
\newblock {Optical chirality and its interaction with matter}.
\newblock {\em Physical Review Letters}, 104(16):1--4, 2010.

\bibitem{Cameron2015}
Robert~P. {Cameron}, J{\"o}rg~B. {G{\"o}tte}, Stephen~M. {Barnett}, and
  Alison~M. {Yao}.
\newblock {Chirality and the angular momentum of light}.
\newblock {\em Philosophical Transactions of the Royal Society of London Series
  A}, 375(2087):20150433, February 2017.

\bibitem{Cameron&Barnett:2012}
Robert~P Cameron and Stephen~M Barnett.
\newblock Electric-magnetic symmetry and {N}oethers theorem.
\newblock {\em New Journal of Physics}, 14(12):123019, dec 2012.

\bibitem{Bliokh_etal:2013}
Konstantin~Y Bliokh, Aleksandr~Y Bekshaev, and Franco Nori.
\newblock Dual electromagnetism: helicity, spin, momentum and angular momentum.
\newblock {\em New Journal of Physics}, 15(3):033026, mar 2013.

\bibitem{Aghapour_etal:2019}
Sajad {Aghapour}, Lars {Andersson}, and Kjell {Rosquist}.
\newblock {The zilch electromagnetic conservation law revisited}.
\newblock {\em Journal of Mathematical Physics}, 61(12):arXiv:1904.08639,
  December 2020.

\bibitem{Olver:1993}
Peter~J. Olver.
\newblock {\em Applications of {L}ie groups to differential equations}, volume
  107 of {\em Graduate Texts in Mathematics}.
\newblock Springer-Verlag, New York, second edition, 1993.

\bibitem{Cameron2012}
R.~P. Cameron and S.~M. Barnett.
\newblock {Electric-magnetic symmetry and Noether's theorem}.
\newblock {\em New Journal of Physics}, 14, 2012.

\bibitem{2015NaPho...9..796B}
K.~Y. {Bliokh}, F.~J. {Rodr{\'\i}guez-Fortu{\~n}o}, F.~{Nori}, and A.~V.
  {Zayats}.
\newblock {Spin-orbit interactions of light}.
\newblock {\em Nature Photonics}, 9(12):796--808, December 2015.

\bibitem{PhysRevA.97.043840}
D.~A. Smirnova, V.~M. Travin, K.~Y. Bliokh, and F.~Nori.
\newblock Relativistic spin-orbit interactions of photons and electrons.
\newblock {\em Phys. Rev. A}, 97:043840, Apr 2018.

\bibitem{DRESSEL20151}
Justin Dressel, Konstantin~Y. Bliokh, and Franco Nori.
\newblock Spacetime algebra as a powerful tool for electromagnetism.
\newblock {\em Physics Reports}, 589:1 -- 71, 2015.

\bibitem{2008NatPh...4..817I}
William T.~M. {Irvine} and Dirk {Bouwmeester}.
\newblock {Linked and knotted beams of light}.
\newblock {\em Nature Physics}, 4(10):817, October 2008.

\bibitem{2013PhRvL.111o0404K}
Hridesh {Kedia}, Iwo {Bialynicki-Birula}, Daniel {Peralta-Salas}, and William
  T.~M. {Irvine}.
\newblock {Tying Knots in Light Fields}.
\newblock {\em Physical Review Letters}, 111(15):150404, October 2013.

\bibitem{2017PhR...667....1A}
M.~{Array{\'a}s}, D.~{Bouwmeester}, and J.~L. {Trueba}.
\newblock {Knots in electromagnetism}.
\newblock {\em Physics Reports}, 667:1--61, January 2017.

\bibitem{2018CQGra..35x5010S}
Tomasz {Smo{\l}ka} and Jacek {Jezierski}.
\newblock {Simple description of generalized electromagnetic and gravitational
  hopfions}.
\newblock {\em Classical and Quantum Gravity}, 35(24):245010, December 2018.

\end{thebibliography}

\end{document}